\documentclass[final]
 {aipproc}

\layoutstyle{6x9}
\newcommand{\fis}{$^{235}U(n_{th},f) $}
\begin{document}

\title[Effects of Neutron Emission on Fragment Mass and Kinetic Energy Distribution ]{\vspace{1.3cm}Effects of Neutron Emission on Fragment Mass and Kinetic Energy
    Distribution from Thermal Neutron-Induced Fission of $^{235}U$}


\author{M. Montoya}{
 address={Instituto Peruano de Energ\'{\i}a Nuclear, Av. Canad\'a 1470, Lima 41, Per\'u.}
  ,altaddress={Facultad de Ciencias, Universidad Nacional de Ingenier\'{\i}a, \\
Av. Tupac Amaru 210, Apartado 31-139, Lima, Per\'u.}
}

\author{J. Rojas}{
 address={Instituto Peruano de Energ\'{\i}a Nuclear, Av. Canad\'a 1470, Lima 41, Per\'u.}
 ,altaddress={Facultad de Ciencias F\'{\i}sicas, Universidad Nacional Mayor de San Marcos,\\
 Av. Venezuela Cdra 34,  Apartado Postal 14-0149, Lima 1, Per\'u.}
 }

\author{E. Saettone}{
 address={Facultad de Ciencias, Universidad Nacional de Ingenier\'{\i}a, \\
Av. Tupac Amaru 210, Apartado 31-139, Lima, Per\'u.}
}

\begin{abstract}
The mass and kinetic energy distribution of nuclear fragments from thermal
neutron-induced fission of \fis ~have been studied using a Monte-Carlo simulation.
Besides reproducing the pronounced broadening in the standard deviation of the
kinetic energy at the final fragment mass number around m = 109, our simulation
also produces a second broadening around \mbox{ $m$ = 125}.
These results are in good agreement with the experimental data obtained by
Belhafaf {\it et al.} and other results on yield of mass. We conclude that the obtained
results are a consequence of the characteristics of the neutron emission, the sharp
variation in the primary fragment kinetic energy and mass yield curves.
We show that because neutron emission is hazardous to make any conclusion on
primary quantities distribution of fragments from experimental results on final
quantities distributions.
\end{abstract}

\maketitle

%%%%%%%%%%%%%%%%%%%%%%%%%%%%%%%%%%%%%%%%%%%%
%% MAINMATTER
%%%%%%%%%%%%%%%%%%%%%%%%%%%%%%%%%%%%%%%%%%%%

\section{Introduction}

Fragment mass and kinetic energy distributions from thermal neutron-induced fission of \fis ~are ones of the most studied parameters since the discovery of the  neutron-induced fission of uranium by Hahn and Strassmann in  1938~\cite{hahn}. The objective was to understand the fission process between the saddle point to scission. Nevertheless, direct measurements can only be  carried out on the final fragments (post neutron emission)  mass yield $Y(m)$ and  kinetic  energy ($e(m)$).

For \fis ~reaction, the mean value of kinetic energy $\overline e$ and the standard deviation  (SD) of  the kinetic energy $\sigma_{e}$ as  function of the final mass  $m$ was measured by Brissot \mbox{ { \it et  al.}~\cite{brissot}}. The plot of the measured $\sigma_{e}$ shows one pronounced broadening around   $m = 109$, which is explained as a results of neutron emission  from nuclear fragments. In a latter experiment, Belhafaf {\it et al.} ~\cite{belha}, repeated the experiment of Brissot { \it et  al.}, obtaining a second broadening around $m = 125$. They claim that this broadening  must exist in the primary fragment kinetic energy ($E(A)$) distribution.

In this  paper, we present a new Monte-Carlo  simulation results concerning \fis. We show that the broadenings on the  $\sigma_{e}$ curve around  the  final fragment  masses  $m = 109$  and $m = 125$ can be reproduced without assuming an adhoc initial structure on $\sigma_{E}(A)$ curve.
 
\section{Monte Carlo simulation model}
\label{sec:model}
%\subsection{Neutron multiplicity as a function of fragment kinetic energy}

In the process of \fis, the excited composed nucleus $^{236}U^*$ is formed first. Then, this nucleus splits in two complementary fragments. Assuming a linear dependence between kinetic energy and number of emitted neutrons, and taking into account that there is no neutron emission ($\nu = 0$) for fragments having the maximal kinetic energy ($E_{max}$)  and that for the average value of fragment kinetic energy ($\bar e$) the neutron number is equal to  $\bar \nu$,  the neutron number $N$  as a function of kinetic energy results,
\begin{equation}
N = {\rm Integer ~ part ~ of}[\alpha + \bar \nu (1 - \beta (\frac{E- \bar E}{\sigma_E}))],
\label{eq:nu}
\end{equation}
where  $\beta$ define the maximal value of kinetic energy as $E_{max} = \bar E + \frac{\sigma_E}{\beta}$,  and  $\alpha$ is used to compensate the effect of the change from a real number $\nu$ to an integer number $N$.

\subsection{Simulation process}

In our  Monte Carlo simulation the input quantities are the primary fragment yield ($Y$), the average kinetic energy ($\bar E$), the SD of the kinetic energy distribution ($\sigma_E$) and the average number of emitted neutron ($\bar \nu $) as a function of primary fragment mass ($A$). The output of the simulation for the final fragment are the yield ($Y$), the SD of the kinetic energy distribution ($\sigma_E$) and the average number of emitted neutron ($\bar \nu $) as a function of final fragment mass $m$.

For the first simulation, we take ($Y$) from Ref. ~\cite{Wagemans}, $\bar \nu$ from experimental results by Nishio {\it et al.} ~\cite{Nishio}, and $\bar E$ from Ref. ~\cite{belha}. The first standard deviation $\sigma_E$ curve is taken without any broadening as function of $A$.  Then, we adjust $Y(A)$, $\nu (A)$, $\bar E(A)$ and $\sigma_E(A)$ in order to get $Y(m)$, $\bar \nu $, $\bar e(m)$, $\sigma_e(m)$ in agreement to experimental data.

In the simulation, for each primary mass $A$, the kinetic energy of the fission fragments is chosen randomly from a Gaussian distribution with mean value $\overline{E}$  and SD $\sigma_{E}$.

For each $E$ value, the simulated number of neutrons $N$ is calculated through  the relation (\ref{eq:nu}). The final mass of the fragment is equal to  $m = A-N$.  Furthermore, assuming that the fragments loose energy only by neutron evaporation and not by gamma emission or any other process, and neglecting the recoil effect due to neutron emission, then the kinetic energy $e(m)$ of the final fragment will be given by
\begin{equation}
\label{eq:ef}
e(m)=(1-\frac {N}{A})E.
\end{equation}
With the assemble of values corresponding to $m$, $e$ and $N$, we calculate $Y(m)$, $\bar e(m)$, $\sigma_e(m)$ and $\nu (m)$.

\begin{figure}
%\centering
\includegraphics[angle=270,width=0.8\textwidth ]{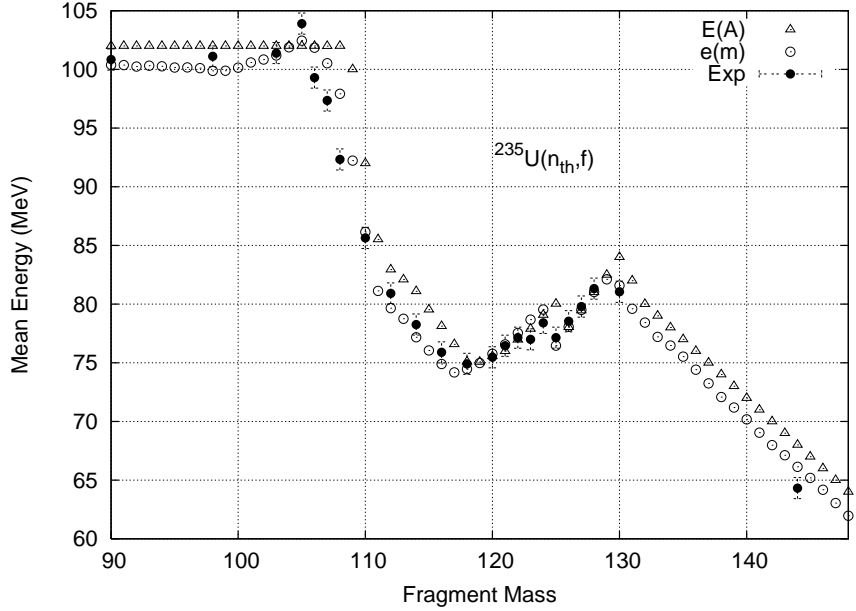}
\caption{Mean kinetic energy of the primary ($\triangle$) and final ($\circ$) fragments, as a result of simulation in this work, to be compared to experimental data ($\bullet$) from Ref. \cite{belha}.}.
\label{fig:ekm}
\end{figure}

\section{Results and discussion}
\label{results}
The plots of the simulated mean kinetic energy for the primary and final fragments as function of their corresponding  masses, are shown in Fig.~\ref{fig:ekm} . In general, the simulated average final kinetic energy curve as a function of final mass ($\bar e(m)$) have roughly a shift similar to that of $Y(m)$ curve, and a diminishing given by relation (\ref{eq:ef}) with $N=\bar \nu$. The exceptions of this rule are produced in mass regions corresponding to variations of the slope of $Y(A)$ or $\bar E(A)$ curves, for example for $A=109$, $A=125$ and $A=130$.

\begin{figure}
%\resizebox{0.8\columnwidth}{!}
\includegraphics[angle=270,width=0.8\textwidth]{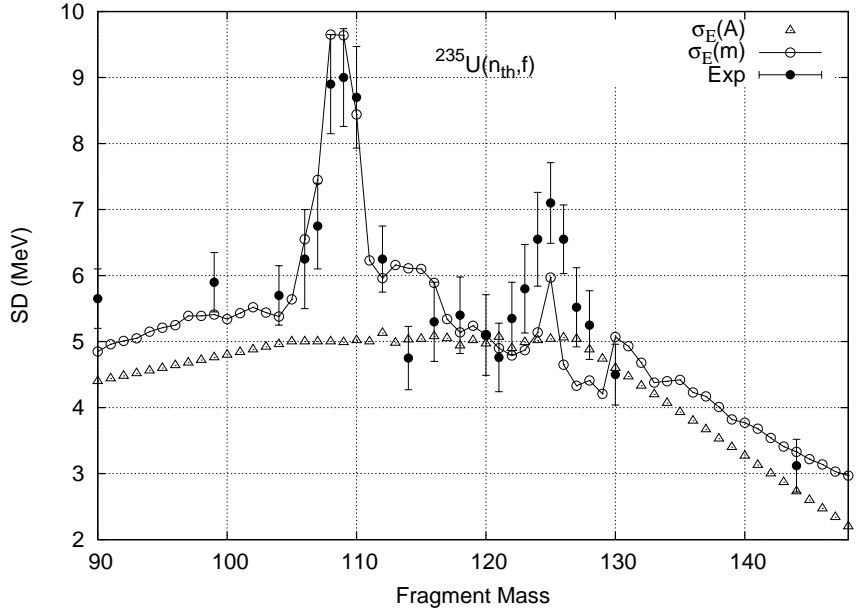}
\caption{SD of primary fragment kinetic energy distribution ($\triangle$) and SD of final fragment kinetic energy distribution ($\circ$), as simulated in this work, to be compared to experimental data ($\bullet$) from Ref. \cite{belha}.}
\label{fig:sde}
\end{figure}

Furthermore, Fig.~\ref{fig:sde} displays  the SD of the kinetic energy distribution of the primary fragments and the SD of the kinetic energy of the final fragments ($\sigma_{e}(m)$). The simulated results for $\sigma_{e}(m)$ presented in Fig. \ref{fig:sde} were obtained  with $\alpha$ = 0.62 and $\beta$=0.35. The plots of $\sigma_{e}(m)$ reveal the presence  of a  pronounced broadening around  $m$ = 109, and a second broadening is found around $m = 125$, in a mass region where there are variations of the slopes of $Y(A)$ or $\bar E(A)$ curves. There is no experimental data around $m=130$. Nevertheless, if one takes the experimental value $\sigma_e=3.9 MeV$ for $m=129$ from Ref. \cite{brissot} and one puts it on Fig. ~\ref{fig:sde}, the beginning of another broadening for $m=130$ is suggested. These results were obtained with a simulated primary fragment kinetic energy distribution without broadenings in the range of fragment masses $A$  from 90 to 145 \mbox{(see Fig. \ref{fig:sde}}, $\triangle$). If one simulates an additional source of energy dispersion in $\sigma_E$, without any broadening, no broadening will be observed on $\sigma_e$.

The presence of broadenings on $\sigma_e$ about $m = 109$ could be associated with neutron emission characteristics (approximately $\bar \nu = 2$) and a very sharp fall in kinetic energy from \mbox{$E$ =100 MeV} to $E$ = 85.5 MeV, corresponding to $A$ = 109  and $A$ = 111, respectively. 
The second broadening is produced by  a discontinuity of the curve $\bar E(A)$ around \mbox{$A$ = 126}, which is necessary to reproduce a similar discontinuity on $e(m)$ around  $m$ = 125. We give emphasis to the shape of $\sigma_e$ which increase from $m = 121$ to $m = 125$ and it decreases from $m = 125$ to $m = 129$ as occurs with experimental data.

\section {Conclusions}
For \fis, in comparison with the primary fragments, the final fission fragments have eroded kinetic energy and mass values, as much as to give rise to the appearance of broadenings in the SD of the final fragments kinetic energy as a function of mass ($\sigma_{e}(m)$) around $m$ = 109 and $m$ = 125 respectively. These broadenings are a consequence of neutron emission and variations on slopes of primary fragments yield ($Y(A)$) and mean kinetic energy $\bar E(A)$ curves. From our simulation results, another broadening, around $m=130$, may be predicted.
%\begin{theacknowledgments}
%\end{theacknowledgments}

\end{document}